\documentstyle[prd,aps,epsfig,rotating]{revtex}
\input epsf
\begin{document}
\onecolumn
\begin{titlepage}
{\flushright      WATPHYS-TH99/02}\\
{\flushright ITP NSF-ITP-99-24}\\
\begin{center}
{\Large \bf Entropy of Rotating Misner String Spacetimes} \\ \vspace{2cm}
R.B. Mann\footnotemark\footnotetext{email: 
rbmann@itp.ucsb.edu \\
on leave from Dept. of Physics, University of Waterloo, Waterloo
Ont. Canada N2L 3G1} \\
\vspace{1cm}
Institute for Theoretical Physics\\
Dept. of Physics\\
University of California Santa Barbara\\
Santa Barbara, CA USA 93106\\
\vspace{2cm}
\vspace{2cm}
\today\\
\end{center}
\begin{abstract}
Using a boundary counterterm prescription motivated by the AdS/CFT conjecture,
I evaluate the energy, entropy and angular momentum 
of the class of Kerr-NUT/bolt-AdS spacetimes. As in the non-rotating case,
when the NUT charge is nonzero the entropy is no longer equal 
to one-quarter of the area due to the presence of the Misner string. 
When the cosmological constant is also non-zero, the entropy is bounded from above. 
\end{abstract}

\end{titlepage}

The thermodynamic properties of gravity have for long appeared to be inextricably 
connected with the presence of black holes \cite{beck,hawk1}. A physical entropy $S$
and temperature $\beta^{-1}$ can be ascribed to a given black hole configuration, where these
quantities are respectively proportional to the area and surface gravity of the
event horizon(s). 

Recently it has been demonstrated that entropy can be associated with a broader
and qualitatively different gravitational system, one containing
Misner strings \cite{Hawkhunt}.  These objects are the gravitational analogues of Dirac strings, and
arise whenever the gravitational field in the Euclidean regime has
a U(1) isometry group (generated by a timelike Killing vector $\xi$)
with a fixed point set of co-dimension $d_f < d-2$ (called a ``nut'' \cite{gibh}).
The existence of any fixed point set makes it impossible to everywhere foliate the spacetime 
with surfaces of constant $\tau$, leading to a difference between the total energy 
($H_{\infty}$) 
and free energy of the gravitational system, which thermodynamically is proportional to its entropy.
In general a spacetime can contain both black holes (for which $d_f=d-2$ -- called a ``bolt'') 
and Misner strings, and the total gravitational entropy will receive contributions from both 
of these objects. 

Explicit demonstration of these ideas has been given in a number of cases, for spacetimes with
and without cosmological constant.  At first it was thought that the presence of a bolt
(i.e. black hole) was necessary \cite{Hawkhunt,HPH,Chamb} since the Misner string contribution
to the entropy is divergent -- only the relative entropy (and energy) between
a spacetime with a bolt/nut configuration and its asymptotically matched pure nut counterpart
was calculated. However it has more recently been shown that
Misner strings themselves have an intrinsic entropy \cite{nutent,EJM}, even
if no bolts are present. By adding to the action
an additional boundary term which is a functional of the intrinsic curvature invariants on
the boundary, the equations of motion are unaffected and the gravitational entropy 
(and total and free energies) is finite whether or not there is a bolt.
The inclusion of this boundary term is motivated from recent work \cite{vjk,hyun} on the
conjectured AdS/CFT duality, which equates the bulk
gravitational action of an asymptotically AdS spacetime with the quantum effective action
of a conformal field theory (CFT) defined on the AdS boundary. The 
coefficients in the additional term may be uniquely fixed by demanding that 
it be finite for Schwarzchild-AdS spacetime. However the spacetime need not be locally
AdS asymptotically -- locally asymptotically flat cases may also be 
included \cite{nutent}.

The purpose of this paper is to extend these considerations to 
to include rotation. Specifically I consider the class of Euclidean Kerr and
Kerr-AdS solutions with and without NUT charge in four dimensions. 
This class of spacetimes forms an important test
case for the counterterm prescription and has received relatively little attention
in the literature \cite{GP,kernut}. Even for Kerr spacetimes with zero nut charge
and cosmological constant the problem of computing quasilocal energy is very difficult 
and has only been
carried out in the slow-rotating limit \cite{Martinez}. The counterterm
prescription given in ref. \cite{nutent} extends to the full Kerr-NUT class, 
reproducing the values of the mass and angular momentum without any background 
spacetime subtractions. When the
nut charge is non-vanishing, I find that the presence of rotation does not admit
the existence of regular spacetime solutions unless a bolt is also present. Furthermore
the rotation parameter has no upper bound when the nut charge and cosmological constant are 
nonzero.    The
entropies for these spacetimes are also computed and are not proportional to their horizon
areas  due to the Misner strings.  This is the first calculation of
the entropy of rotating spacetimes with nut charge. Only the main results will
be presented here; details will appear in a forthcoming paper \cite{forth}.

Consider a Euclidean manifold $M$ with metric $g_{\mu\nu}$,
covariant derivative $\nabla_\mu$, and time coordinate $\tau$ which 
foliates $M$ into non-singular hypersurfaces $\Sigma_\tau$ with unit 
normal $u_{\mu}$. $\Theta^{\mu\nu}$  (whose trace is $\Theta$)
denotes the extrinsic curvature  of any boundary(ies)  $\partial M$ of
the manifold $M$ (internal and/or at
infinity), with induced metric(s) $\gamma$. The path-integral formulation of 
quantum gravity implies that
the Euclidean action $I = -\log Z$ to lowest order in $\hbar$,
where $Z$ is the partition function of an ensemble
\begin{equation}\label{e0}
Z = \int[Dg][D\Phi]\exp[-I(g,\Phi)]
\end{equation}
with the path integral taken over all metrics $g$ and matter fields $\Phi$ that are 
appropriately indentified under the period $\beta$ of $\tau$. The thermodynamic
definition $\log Z = S - \beta H_\infty$ then implies that the entropy of a given
spacetime is
\begin{equation}\label{e1}
S = \beta H_{\infty} - I = \beta(E+\Omega\cdot J)-I
\end{equation}
where $E$ and $J$ are respectively the energy and angular momentum of the spacetime 
at infinity and $\Omega$ the angular velocity at the event horizon. 
The entropy is then the difference between the value 
the action would have ($\beta H_\infty$, the total energy) 
if there were no breakdown of foliation and its  actual value (proportional to the free energy). 

One could compute this difference by removing small neighbourhoods 
$N_\epsilon^i$ of the fixed point sets and strings so that 
$I= I_{M_\epsilon^i}-\sum_i I_{N_\epsilon^i}$. Rewriting the $I_{M_\epsilon^i}$ into
Hamiltonian form (taking care to include the additional surface terms
due to these new boundaries), one finds that the  only non-zero contributions 
to the Hamiltonian are from the boundaries at infinity and along the strings. When the 
contributions $I_{N_\epsilon^i}$ from the small
neighbourhoods of the fixed point sets are re-inserted, their surface terms are 
non-vanishing and yield the one-quarter of the areas of the neighbourhoods 
removed, i.e. of the bolts and the strings.  

The action is generally taken to be a linear combination
of a volume (or bulk) term
\begin{equation}\label{e3}
I_v = -\frac{1}{16\pi}\int_M d^dx\sqrt{g}\left(R+2\Lambda+{\cal L}(\Phi)\right)
\end{equation}
and a boundary term
\begin{equation}\label{e4}
I_b = -\frac{1}{8\pi}\int_{\partial M} d^{d-1}x\sqrt{\gamma}\Theta(\gamma)
\end{equation}
(chosen to yield a well-defined variational principle), where
${\cal L}(\Phi)$ is the matter Lagrangian and $\Lambda$ the cosmological constant.
When evaluated on solutions both 
$I_v$ and $I_b$ are typically divergent, yielding divergent values for both the
string area and Hamiltonian terms and hence for the entropy. One method of dealing
with this difficulty is to compute everything relative to some chosen 
reference background spacetime (suitably matched in its asymptotic and topological
properties) whose boundary(ies) have the same induced metric(s) as those 
in the original spacetime \cite{BY,BCM,HHor}; the reference spacetime is then interpreted
as the vacuum for that sector of the quantum theory. Such a choice is not always 
unique \cite{CCM}, nor is it always possible to embed a boundary with a given induced 
metric into the reference background. Indeed, for Kerr spacetimes this latter problem forms 
a serious obstruction towards calculating
the subtraction energy, and calcuations have only been performed in the slow-rotating 
regime \cite{Martinez}.

The counterterm proposal involves adding a term $I_{ct}$ to the action, where \cite{nutent}
\begin{equation}\label{e7}
I_{ct} = \frac{2}{\ell}\frac{1}{8\pi} \int_{\partial M_\infty} 
d^{3}x\sqrt{\gamma}\sqrt{1 + \frac{\ell^2}{2} R(\gamma)} 
\end{equation}
with $\ell = \sqrt{3/|\Lambda|}$. The coefficients of the $R(\gamma)$ term and 
the overall action are determined
by demanding that the Schwarzchild-AdS solution have finite total action
$I_T=I_v + I_b + I_{ct}$.  The prescription (\ref{e7}) has been shown to be 
sufficient for evaluating
the actions, entropies and total energies for the Schwarzchild, Taub-bolt, and
Taub-NUT spacetimes, along with their AdS and topological extensions 
without the use of any background subtractions \cite{nutent,EJM}. It is motivated by
the conjectured AdS/CFT correspondence: divergences appearing in the stress-energy tensor
of the boundary CFT are just the standard ultraviolet divergences of quantum field theory and may
be removed by adding counterterms to the action which depend only on the 
intrinsic geometry of the boundary. Quantities such as energy, entropy and (as will be shown)
angular momentum are then intrinsically defined for a given spacetime, rather than
with respect to a reference background. Furthermore,
(\ref{e7}) applies even in the $\ell\to \infty$ limit, thereby including
asymptotically locally flat cases, unlike the prescriptions in refs. \cite{EJM,vjk,hyun}
to which (\ref{e7}) reduces for small $\ell$.

Using (\ref{e7}) the entropy is
\begin{equation}\label{e8}
S =  \beta H_\infty - (I_v + I_b + I_{ct}) 
\end{equation}
where all quantities are evaluated on a given solution, and where
$H_\infty = M + \Omega J$, with $M = Q[\partial/\partial\tau]$ and
$J=Q[\partial/\partial\phi]$ being the conserved charges associated
with the Killing vectors $\partial/\partial\tau$ and 
$\partial/\partial\phi$ where $\Omega= a/(r_+^2-a^2-N^2)$, with $r_+$ defined below. 
These conserved charges are given by \cite{BCM,ivan}
\begin{equation}\label{e9}
Q[\xi] = \frac{1}{8\pi}\int_{\partial M_\infty \cap \Sigma_\tau}
\left[\Theta^{\mu\nu}-\Theta \gamma^{\mu\nu} 
+ \frac{2}{\sqrt{\gamma}}\frac{\delta I_{ct}}{\gamma_{\mu\nu}}\right]u_{\mu}\xi_{\nu}
\end{equation}
which may be shown by taking the variation of the action with
respect to the boundary metric $\gamma_{\mu\nu}$ at infinity .

The class of Euclidean Kerr-NUT-AdS spacetimes has the metric form
\begin{eqnarray}
ds^2 &=& 
\frac{V(r)\,({d{\tau }} 
  -(2N\cos(\theta)-a\sin^2(\theta))d\phi)^{2}) 
+ {{\cal H}}(\theta )\,{\rm sin}(\theta )^{2}\,(a\,{d{\tau }}
 - (r^{2} - N^{2} - a^{2})\,{d{\phi }})^{2} 
}{ 
\chi ^{4}\,(r^{2} - (N + a\,{\rm cos}(\theta ))^{2})}\nonumber \\
&&+(r^{2} - (N + a\,{\rm cos}(\theta ))^{2})\,(
{\displaystyle \frac {{d{r}}^{2}}{V(r)} }  + 
{\displaystyle \frac {{d{\theta }}^{2}}{{\rm {\cal H}}(\theta )}} ) 
\label{e9a}
\end{eqnarray}
where the Einstein field equations imply 
\begin{eqnarray}\label{e10}
{\cal H} &=& 1 + {\displaystyle \frac {q\,N^{2}}{l^{2}}}  + {\displaystyle 
\frac{(2\,N + a\,{\rm cos}(\theta ))^{2}}{l^{2}}} \\
\qquad
V(r) &=& {\displaystyle \frac {r^{4}}{l^{2}}}  + {\displaystyle 
\frac{((q - 2)N^{2} -a^2 + l^{2})\,r^{2}}{l^{2}}}  - 2\,m\,r - 
{\displaystyle \frac {(a + N)\,(a - N)\,(q\,N^{2} + l^{2} + N^{2})}{l^{2}}} 
\nonumber
\end{eqnarray}
where the periodicity in $\tau$ and the parameters $q$ and $\chi$ are chosen so that 
conical singularities are avoided. In the $(\theta,\phi)$ section these considerations
imply that $q=-4$ and $\chi=1/\sqrt{1+a^2/l^2}$.

The periodicity in $\tau$ is more subtle.
The location of the nut is at $r=\sqrt{a^2+N^2}\equiv r_N$, where the 
area of surfaces orthogonal to the $(r,\tau)$
section vanishes.  The Misner string singularity runs along the z-axis from the nut
to infinity. Regularity along the z-axis then implies that $\tau$ has period 
$8\pi N$. However regularity in the $(r,\tau)$ section implies that $\tau$ also 
has period $2\pi/\kappa$ where 
\begin{equation}\label{e11}
\kappa = \frac{V^\prime(r_+)}{4\pi\chi^2(r_+^2-r_N^2)}
\end{equation}
where $V(r_+)$=0, $r_+$ being the location of the foliation breakdown. 
Interpreting this latter equation as determining $m$ in terms of $r_+$,
equating these two periods yields a quartic  constraint (cubic if $\l \to \infty$)
on $r_+$ in terms of $a$, $N$  and $l$. A more careful treatment is required if 
$r_+=r_N$; in this case regularity of the solutions 
demands that $V(r)$ have a double root there. However this requirement turns out to be incompatible with the
periodicity constraints unless $a=0$. 
Hence $r_+>r_N$, and there are no regular Kerr-NUT 
or Kerr-NUT-AdS solutions (the former observation was made in ref. \cite{GP}).

Using the formulae (\ref{e9}), I find after somewhat lengthy and tedious calculation 
\begin{equation}\label{e14}
M = \frac{m}{\chi^4}  \qquad  J= \frac{ma}{\chi^4}
\end{equation}
for each of the Kerr, Kerr-AdS, Kerr-bolt and Kerr-bolt-AdS solutions, 
the parameter $m$ obeying the constraints mentioned in the previous paragraph 
in the bolt case.  The actions, Hamiltonians, and entropies
for each case are finite. The results are given in table I; omitted is the 
Hamiltonian for each case, which is simply 
$H_\infty = \frac{m}{\chi^4}(1+a\Omega)$. 
After substituing the solutions for $m$ and $r_+$, the action in the Kerr-bolt
case is $4\pi Nm$, in agreement with ref. \cite{GP}.
Note that for the bolt solutions the entropy is not one-quarter 
of the area, due to the presence of the Misner string.

One of the more unusual results apparent from table I is that in the Kerr-bolt-AdS solution 
the entropy is not positive for all values of the parameters. 
The entropy is always positive for any values of $a$ and $l$ provided
$19 K^6 - 78 K^4+3 K^2 + 8 > 0$, or $K>K_m=2.02282556...$, where $K=r_+/r_N$. However for
$1<K<K_m$ there exist ranges of values of $a/N$ for which the entropy is negative, and
for certain values of $a$ the entropy diverges to $-\infty$. 
Similar properties have been noted in the non-rotating AdS NUT and bolt 
solutions \cite{nutent,EJM}.  The entropy $s=S/N^2$ as a function of $x=a/N$ is plotted 
in Fig.1 for the Kerr-bolt solution -- for small $x$ $s\approx 5\pi+{\cal O}(x^2)$, and for
large $x$, $s \to 4\pi x$. The parameter $m$ is an everywhere defined increasing function
of  $x$, and approaches $2N$ for large $x$. The behaviour is quite different in the  
the Kerr-bolt-AdS solution. For a given $K>K_m$, 
$s$ reaches a maximum somewhere between $x=0$ and $x=x_m$; for $x>x_m$, the parameter $l^2<0$ and so
there are no allowed solutions. For $K<K_m$, the entropy (and $m$) will be negative in some allowed region
of $x$ (i.e. where $l^2>0$. Figures 2 and 3 show typical cases. 

To summarize, the prescription (\ref{e7}) has been shown to apply to spacetimes
with non-zero angular momentum,  and so removes the troublesome aspects of 
evaluating physical quantities in gravity relative to some chosen background
\cite{Martinez,CCM}. Indeed, the expressions for $E$, $J$ and $S$ could all be given 
quasilocally at finite radius $R$, although I have omitted them here for the sake of brevity.
As with the non-rotating case, these results must be carefully interpreted.
For $N\neq 0$ the boundary at infinity is not a direct product $S^1\times S^2$
but instead is a squashed $S^3$.  Consequently continuation to the Lorentzian regime is 
not straightfoward the way it is in the $N=0$ cases. The most promising possibility 
is that of interpreting the path integral over all metrics as the partition function for
an ensemble of spacetimes with fixed NUT charge \cite{Hawkhunt,HPH}.  A more complete understanding
of the thermodynamics of these solutions (and how to interpret the negative values 
of the entropy in the AdS case), as well as the relationship between these results and
the behaviour of a conformal field theory on the boundary remain interesting questions
for further study.

\begin{figure}
\begin{center}
\epsfig{file=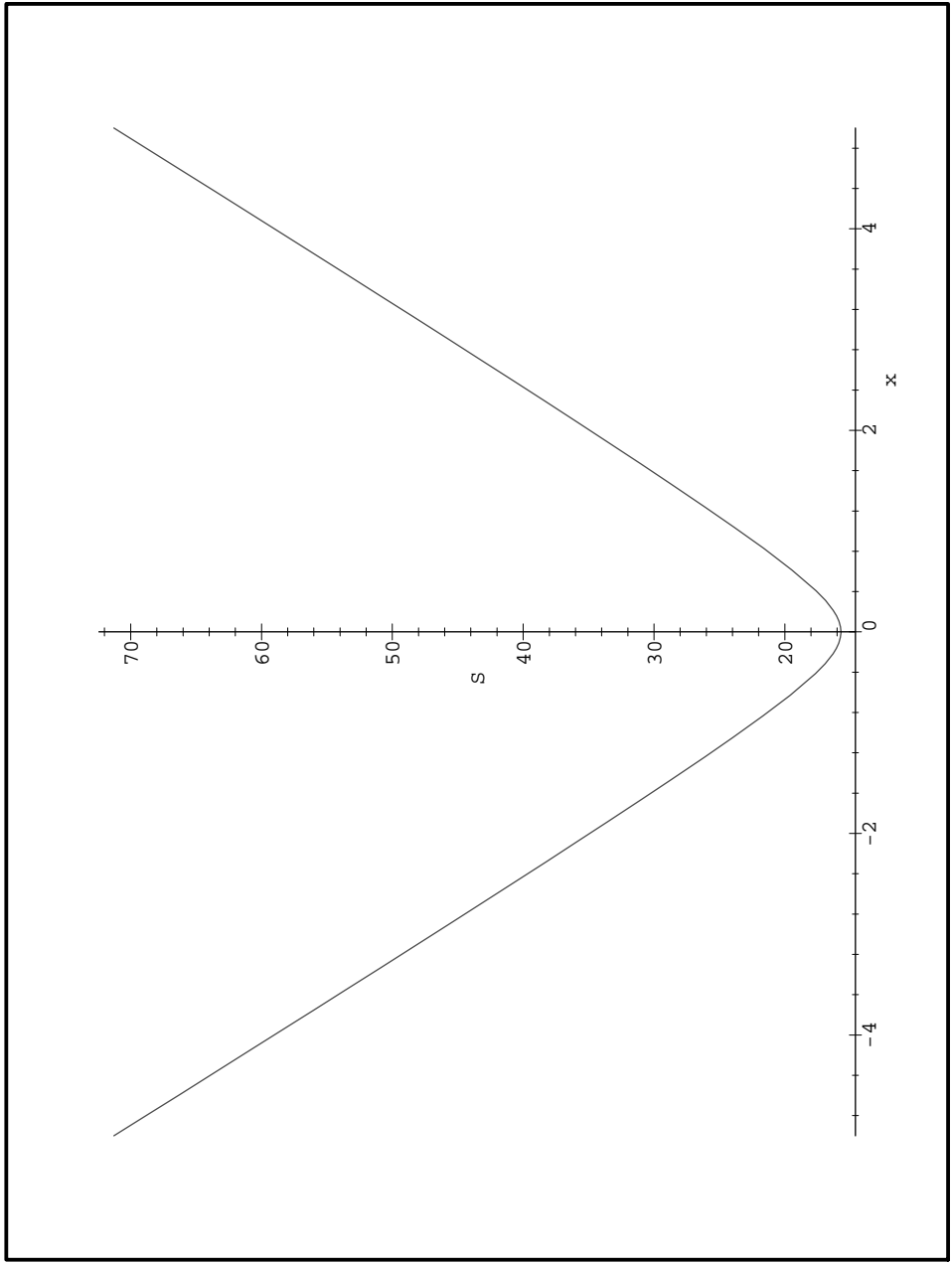,width=0.8\linewidth}
\end{center}
\caption{The entropy as a function of $x=a/N$ in the Kerr-bolt case}
\label{fig1}
\end{figure}

\begin{figure}
\begin{center}
\epsfig{file=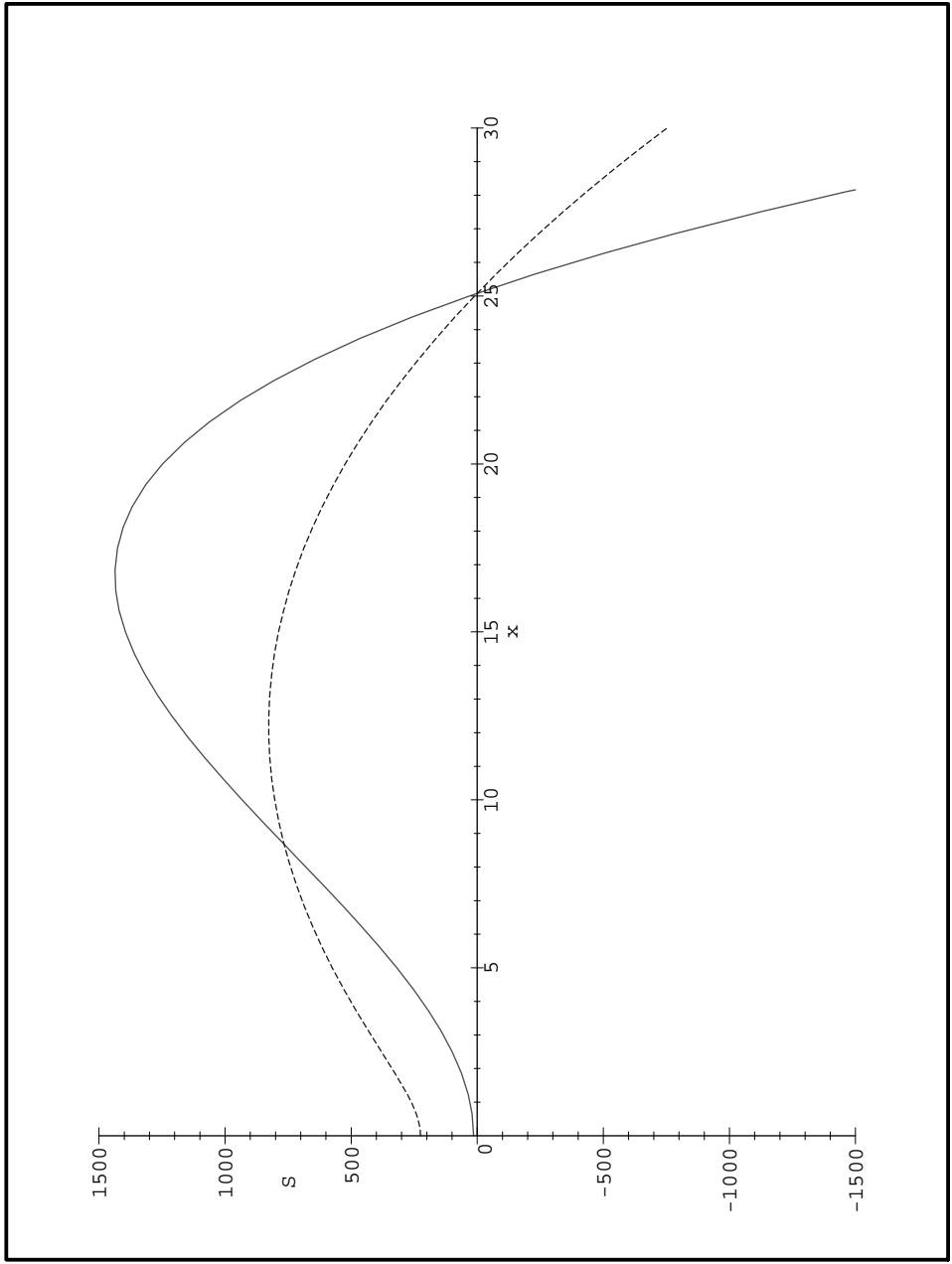,width=0.8\linewidth}
\end{center}
\caption{The entropy (solid line) and $l^2/N^2$ (dashed line) 
as a function of $x=a/N$ for $K=4$ in the Kerr-bolt-AdS case.}
\label{fig2}
\end{figure}

\begin{figure}
\begin{center}
\epsfig{file=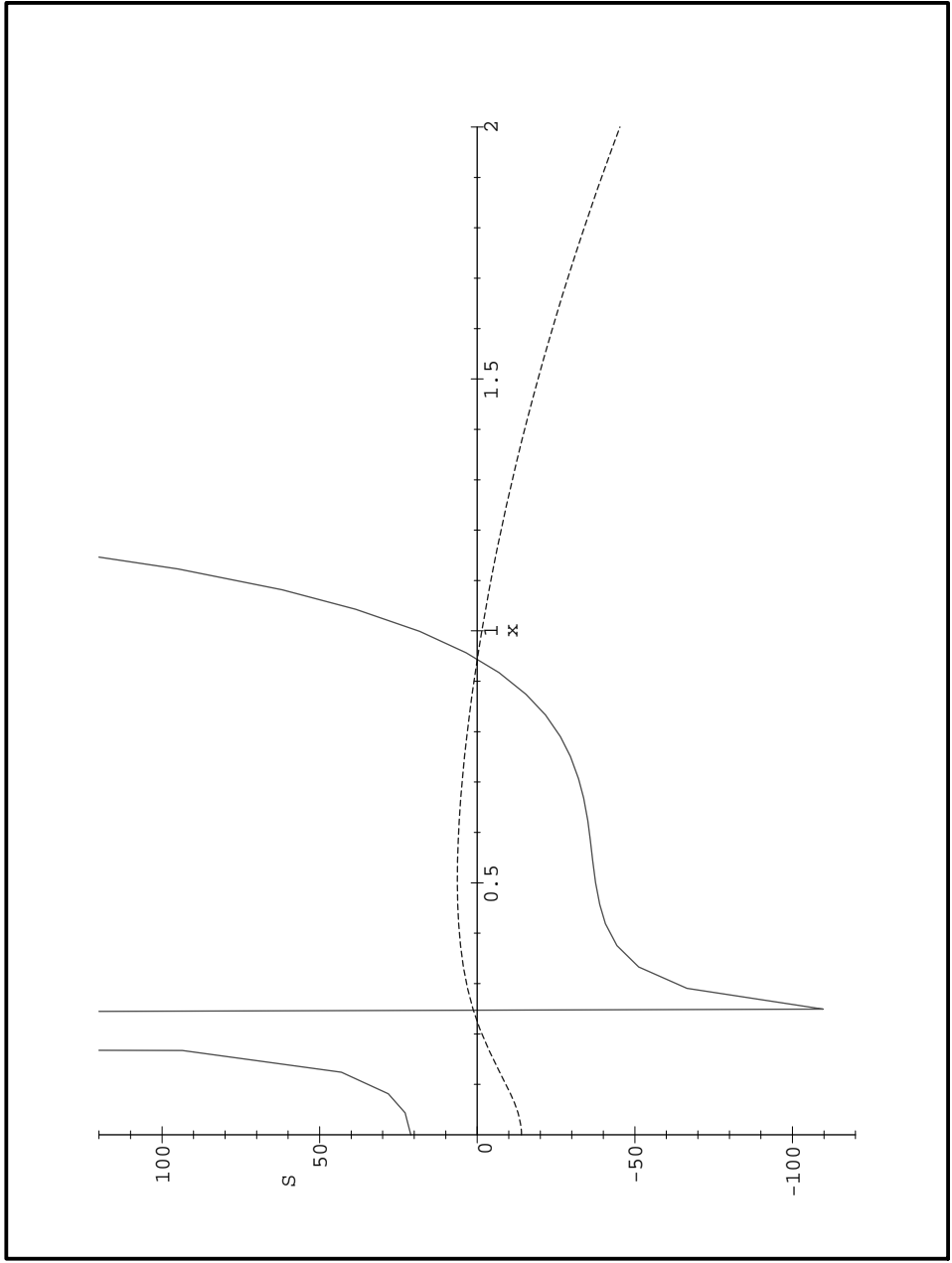,width=0.8\linewidth}
\end{center}
\caption{The same plot as Fig. 2, but for $K=1.01$.}
\label{fig3}
\end{figure}

\vskip 0.3 in
The work was supported by a National Science Foundation grant PHY94-07194 and 
by the Natural Sciences and Engineering Research Council of Canada. 
I am grateful to the ITP and to the physics department at UCSB for their hospitality.

\onecolumn
\begin{tabular}{|c|c|c|c| }
\multicolumn{4}{c}{\Large Table I}\\
\hline 
 & & &  \\
Spacetime & Periodicity & Action & Entropy \\ 
 & & & \\ \hline
 & & &  \\ 
Kerr &  $\frac{4\pi r_+(r_+^2-a^2)}{(r_+^2+a^2)}$  & $\frac{\pi(r_+^2-a^2)^2}{(r_+^2+a^2)}$ 
   & $\pi(r_+^2-a^2)$ \\
             & & & \\ \hline  
& & &  \\
Kerr-bolt &  $8\pi N$ & $\pi\frac{(r_+^2-a^2)^2-N^4}{r_+^2+a^2-N^2}$ & $\pi(r_+^2-a^2+N^2)$ \\
             & & &  \\ \hline  
& & &  \\
Kerr-AdS &  $\frac{4\pi r_+(\ell^2+a^2)(r_+^2-a^2)}{3r_+^4+(\ell^2-a^2)r_+^2+a^2 \ell^2}$  & 
$-\frac{ \pi (r_+^2-\ell^2) (r_+^2-a^2)^2}{(3r_+^4+(\ell^2-a^2)r_+^2+a^2 \ell^2) \chi^2}$ 
& $\pi\frac{r_+^2-a^2}{\chi^2}$ \\
             & & &  \\ \hline 
& & &  \\
Kerr-bolt-AdS & $8\pi N$ &

$-\pi\frac{(r_+^2-N^2-a^2)[r_+^4-(a^2+\ell^2)   r_+^2 + (N^2-a^2) (3N^2-\ell^2)]}
     {(3r_+^4+(\ell^2-a^2-6N^2)r_+^2+(N^2-a^2)(3N^2-\ell^2))\chi^2}$
& 
$\pi\frac{3r_+^6+(\ell^2-4a^2-15N^2)r_+^4+(a^2+3N^2)^2 r_+^2+(N^2-a^2)^2 (3N^2-\ell^2)}
     {(3r_+^4+(\ell^2-a^2-6N^2)r_+^2+(N^2-a^2)(3N^2-\ell^2))\chi^2}$
 \\
             & & &  \\ \hline  
\end{tabular}

\end{document}